\documentclass[pre,floats,twocolumn,superscriptaddress,usenames,dvipsnames,nofootinbib]{revtex4-1}

\usepackage{graphicx}
\usepackage{rotate}
\usepackage{epsfig}
\usepackage{xcolor}  
\usepackage[normalem]{ulem}
\usepackage{amssymb,amsfonts,amsmath}
\usepackage{url}
\usepackage{multirow}
\usepackage{dcolumn} 
\usepackage{paralist}

\newcommand{\avg}[1]{\langle #1 \rangle}

\newcommand{\etal}{\emph{et al.} } 
\newcommand{\ie}{\emph{i.e.}, }
\newcommand{\eg}{\emph{e.g.}, }
\newcommand{\Eint}{E_{\text{int}}}
\newcommand{\Eloc}{E_{\text{loc}}}
\newcommand{\Eglob}{E_{\text{glob}}}
\newcommand{\Ltot}{L_{\text{tot}}}
\newcommand{\Lcg}{L_{\text{cg}}}
\newcommand{\Lnorm}{L^{\prime }}

\graphicspath{{pictures/}}

\begin{document}

\title{Comparing spatial networks: A 'one size fits all' efficiency-driven approach}

\author{Ignacio Morer}
\email{ignacio.morer@gmail.com}
\affiliation{Departament de Fisica de la Mat\`eria Condensada, Universitat de Barcelona, Barcelona, Spain}
\affiliation{Universitat de Barcelona Institute of Complex Systems (UBICS) Universitat de Barcelona, Barcelona, Spain}

\author{Alessio Cardillo}
\email{alessio.cardillo@urv.cat}
\affiliation{Institut Catal\`a de Paleoecologia Humana i Evoluci\'o Social (IPHES), E-43007 Tarragona, Spain}
\affiliation{Department of Engineering Mathematics, University of Bristol, Bristol, BS8 1UB, United Kingdom}
\affiliation{Department of Computer Science and Mathematics, Universitat Rovira i Virgili, E-43007 Tarragona, Spain}
\affiliation{GOTHAM Lab -- Institute for Biocomputation and Physics of Complex Systems (BIFI), Universidad de Zaragoza, E-50018 Zaragoza, Spain}

\author{Albert D\'iaz-Guilera}
\email{albert.diaz@ub.edu}
\affiliation{Departament de Fisica de la Mat\`eria Condensada, Universitat de Barcelona, Barcelona, Spain}
\affiliation{Universitat de Barcelona Institute of Complex Systems (UBICS) Universitat de Barcelona, Barcelona, Spain}

\author{Luce Prignano}
\email{luceprignano@ub.edu}
\affiliation{Departament de Fisica de la Mat\`eria Condensada, Universitat de Barcelona, Barcelona, Spain}
\affiliation{Universitat de Barcelona Institute of Complex Systems (UBICS) Universitat de Barcelona, Barcelona, Spain}

\author{Sergi Lozano}
\email{slozanop@ub.edu}
\affiliation{Universitat de Barcelona Institute of Complex Systems (UBICS) Universitat de Barcelona, Barcelona, Spain}
\affiliation{Institut Catal\`a de Paleoecologia Humana i Evoluci\'o Social (IPHES), E-43007 Tarragona, Spain}
\affiliation{\`Area de Prehist\`oria, Universitat Rovira i Virgili, Tarragona, Spain}
\affiliation{Departament d'Hist\`oria Econ\`omica, Institucions, Pol\'itica i Economia Mundial, Universitat de Barcelona, Barcelona, Spain}

\begin{abstract}
Spatial networks are a powerful framework for studying a large variety of systems belonging to a broad diversity of contexts: from transportation to biology, from epidemiology to communications, and migrations, to cite a few. Spatial networks can be described in terms of their total cost (\ie the total amount of resources needed for building or traveling their connections). Here, we address the issue of how to gauge and compare the quality of spatial network designs (\ie efficiency vs. total cost) by proposing a two-step methodology.  Firstly, we assess the network's design by introducing a quality function based on the concept of network's efficiency. Second, we propose an algorithm to estimate computationally the upper bound of our quality function for a given network. Complementarily, we provide a universal expression to obtain an approximated upper bound to any spatial network, regardless of its size. Smaller differences between the upper bound and the empirical value correspond to better designs. Finally, we test the applicability of this analytic tool-set on spatial network data-sets of different nature.

\end{abstract}

\maketitle


\section{Introduction}
\label{sec:intro}

A large variety of systems, both natural and artificial, are composed by interconnected units embedded in space. All these systems can be mapped onto spatial networks \cite{hagget-chorley-1969, barthelemy-phys_rep-2011}, a powerful framework which provides the mathematical and conceptual tools to study them formally. Such a framework, built on basic common features, applies to a broad diversity of contexts: from transportation \cite{guimera-pnas-2005,gastner-jstat-2006,porta-epb-2006,kaluza-jrosocint-2010,rinaldo-pnas-2014} to biology \cite{lui-jchemphys-2013, sandhu-cell_rep-2012}, from neuroscience \cite{bullmore-nat_rev_neu-2012} to animal behavior \cite{perna-jrosocint-2014,bottinelli-jrosocint-2015,cook-etal-2014}, from epidemiology \cite{isella-jtheobio-2010, balcan-pnas-2009, poletto-jtheobio-2013}  to communications \cite{blondel-epjds-2015} and migrations \cite{porat-epl-2016}, to cite a few. 

Spatial networks are those whose nodes have associated spatial coordinates. Consequently, links -- that is, connections between nodes -- are characterized by the distance between the pair of nodes they connect. Such a distance usually, translates into a \emph{cost}, standing for the amount of resources needed for building or traveling (or both) a given connection. For example, the cost to build a road connecting two cities increases as its extremities are farther from each other.

Most of the literature on the topic (\cite{gastner-jstat-2006,gastner-epjb-2006,louf-pnas-2013}) considers the case of edges' costs directly proportional to their length (\ie distances between connected nodes). Despite other options are possible (\eg cost proportional to a monotonically increasing function of the connection's length), this is generally considered a good description of almost the totality of the systems representable as spatial networks. Notice that construction or operative costs can also be assigned to nodes in particular circumstances (\eg routers in communication networks). In this article, we neglect such node costs. By doing this, spatial networks can be characterized in terms of a \emph{total cost} defined as the sum of their edges' costs.

When comparing two spatial networks with the same node layout it is reasonable to think that the one with a higher total link length can achieve better efficiency. Likewise, the same total link length might not be able to communicate in an equally efficient way two sets of nodes, that are different in number and relative location. Our goal is to enable a meaningful comparison between two spatial networks even in such cases. To reach that goal, we assume the total cost to be an external constraint (\ie determined by external factors fixing the amount of resources available for building connections), and focus on assessing to what extent such resources have been employed profitably. Specifically, we frame the quality of a network with regards to the total cost and establish a reference of how better could it be under that constraint, similarly to what was done in \cite{zamora-commphys-2019} for non-spatial networks. It is also a very similar approach to that of Cardillo \etal in which real networks (urban street patterns) are compared by rescaling both their efficiencies and costs with values obtained from suitable null-cases \cite{cardillo-pre-2006}. 

Here, we propose a two-step methodology: \begin{inparaenum}[(1)] \item to assess the design of different networks by means of a quality function; \item to compare those values with an upper bound estimated computationally and therefore calculate how far each system lies from an optimal, equally constrained network.\end{inparaenum}

The paper is organized as follows. After characterizing the behavior of some reference models of spatial networks, we introduce the concept of \emph{integrated efficiency}, $\Eint $, and set it as our quality function  (Sec.~\ref{sec:efficiency}). Then, in Sec.~\ref{sec:assessing} we devise an algorithm to estimate the maximum value that such a metric can take (upper bound) for a given set of node positions (node layout) and a given total link length (constrained total cost). We then provide an approximate universal relation that allows to determine the upper bound of the integrated efficiency as a function of the average distance between nodes and the total link length, for any number of nodes. In Sec.~\ref{sec:application}, we measure the quality of several empirical systems and put them in context with their optimal counterparts, to enable a meaningful comparison regardless of their individual characteristics.  Finally, as an example, we present some potential applications.

\section{Global, Local, and Integrated Efficiencies}
\label{sec:efficiency}

Given a spatial network $G$ with $N$ nodes, its structure is completely determined by the adjacency and distance matrices, $\mathcal{A}$ and $\mathcal{D}$. Their corresponding elements $\{ a_{ij}\}$ take value $1$ $(0)$ if the connection exists (does not exist), and $\{ d_{ij}\}$ take finite positive values corresponding to the spatial distance between nodes $i$ and $j$ \cite{latora-book-2017}. These two matrices fully determine the shortest paths matrix $\mathcal{L}$, whose elements $\{ l_{ij}\}$ stand for the length of the shortest path between nodes $i$ and $j$. $l_{ij}$ is a straight sum of the weights of the links in the path, no matter the number of steps\footnote{It may happen that a path including more links is shorter than another with fewer but longer steps.}.

 There are several ways to assess the quality of a spatial network. We chose to use the concept of \emph{network efficiency} which is, essentially, a comparison between the spatial distance and the shortest path. The efficiency in the communication between two nodes $i$ and $j$, $E_{ij}$ corresponds to the ratio between the spatial distance and the shortest path length, $E_{ij} = d_{ij}/l_{ij}$, commonly known  as \emph{detour index} or \emph{route factor} \cite{barthelemy-phys_rep-2011}. By computing the average over all pairs of nodes we obtain the \emph{global efficiency} as proposed in \cite{vragovic-pre-2005}, (despite an alternative definition has been given in \cite{latora-marchiori-2001}):
\begin{equation}
\label{eq:global_eff}
\Eglob = \frac{1}{N\left( N-1\right)}  \sum_{i\neq j} \frac{d_{ij}}{l_{ij}} \,.
\end{equation}
$\Eglob$ quantifies the ability of the system as a whole to communicate efficiently among its elements. Additionally, it is also relevant to assess the fault tolerance of the system's communicability at local level. To this aim, Latora and Marchiori \cite{latora-marchiori-2001} introduced the so-called \emph{local efficiency}, $\Eloc$. Such an indicator measures how efficient is the communication in the local neighborhood of a node $i$ after its removal, thus accounting for the robustness of the connectivity against local damage.. In this paper, we adopt a modified $\Eloc$ proposed by Vragovic \etal  \cite{vragovic-pre-2005} that measures the efficiency of the communication between any two neighbors $j$ and $m$ of node $i$ -- no matter how far they are -- considering all possible paths connecting them:
\begin{equation}
\label{eq:local_eff}
\Eloc = \frac{1}{N} \sum_{i=1}^N \frac{1}{k_i\left( k_i-1\right)} \sum_{j\neq m \in \Gamma_i} \frac{d_{jm}}{l_{jm/i}} \,,
\end{equation}
where $\Gamma_i$ represents the local sub-graph of neighbors of node $i$ and $l_{jm/i}$ is the length of the shortest path joining nodes $j$ and $m$ in absence of $i$, and $l_{jm/i} = \infty$ when $j$ and $m$ belong to disconnected components. Finally, $k_i$ is the degree (\ie number of connections) of node $i$. 

Given a certain layout of the nodes in space, there are multiple connectivity patterns presenting (approximately) the same \emph{total length}, $\Ltot = \sum_{i,j} a_{ij} \, d_{ij}$. Among the plethora of spatial network models available in the literature \cite{gastner-jstat-2006,gastner-epjb-2006,banavar-prl-2000,kaiser-pre-2004,barthelemy-jstat-2006,barthelemy-prl-2008}, we selected a few models as benchmarks, thus encompassing a wide spectrum of possibilities:
\begin{itemize}
    \item the Minimum Spanning Tree (MST) \cite{barthelemy-phys_rep-2011};
    \item the Greedy Triangulation (GT), a maximally connected  triangulation minimizing its total length \cite{cardillo-pre-2006};
    \item the Equitable Efficiency Model (EEM), a growth model adding one link at a time ensuring that such move constitutes the best improvement in the communication among any pair of nodes \cite{prignano-jas-2019};
    \item the Gastner-Newman model (GN), a growth model where edges have an \emph{effective length} which combines Euclidean and topological distances altogether \cite{gastner-epjb-2006}.
\end{itemize}
A detailed explanation of all the models (except for the well-known Minimum Spanning Tree) can be found in the Appendix.

\begin{figure}
\centering
\includegraphics[width=\columnwidth]{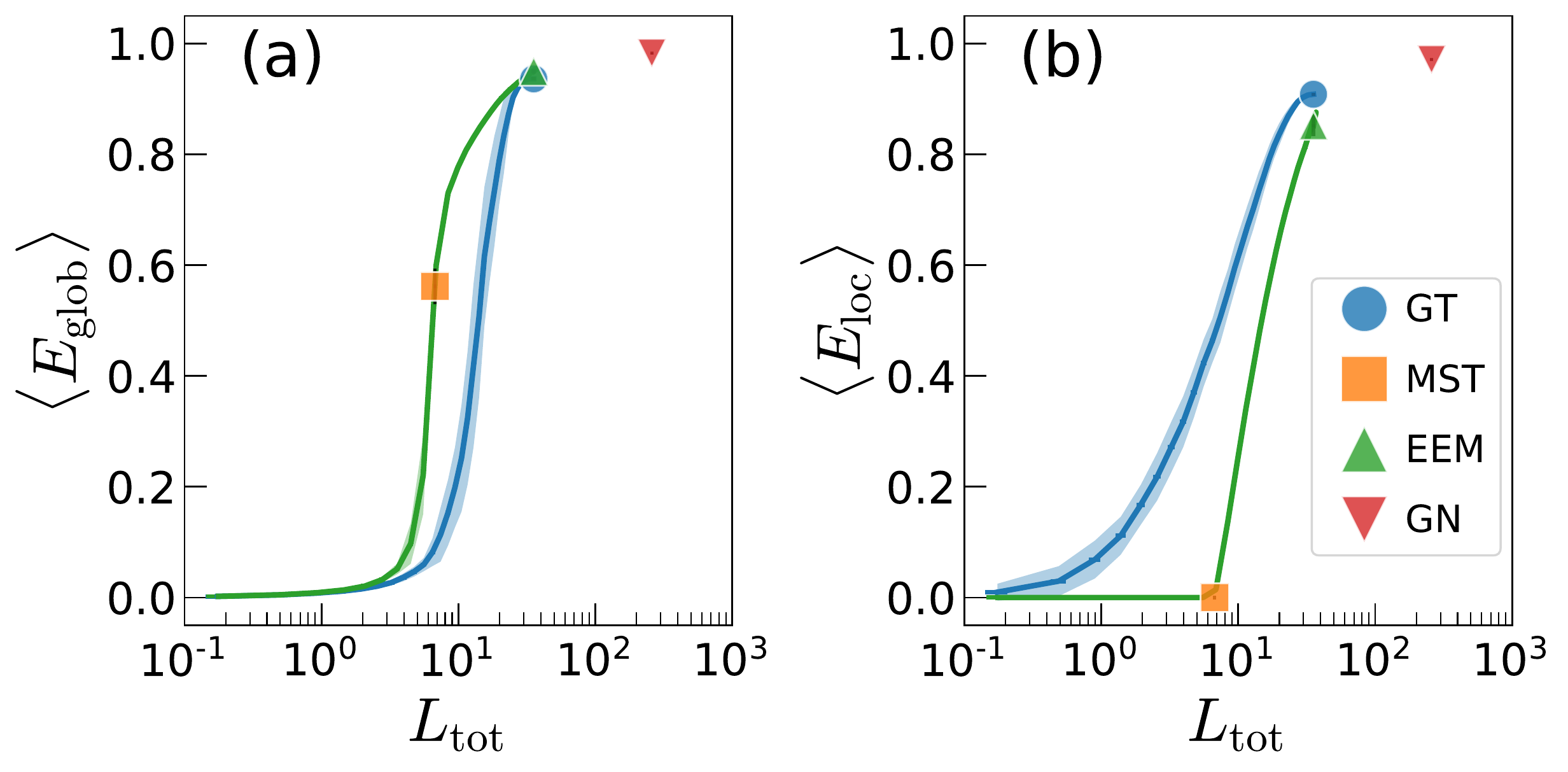}
\caption{Average global  efficiency, $\avg{\Eglob}$, (panel a), and average local  efficiency, $\avg{\Eloc}$, (panel b) as a function of the total length of the system, $\Ltot$, for different models: Greedy Triangulation (GT),  Minimum Spanning Tree (MST), Equitable Efficiency Model (EEM), and Gastner-Newman (GN) with $\lambda = 0.85$ and $\gamma = 0.06$. Dots refer to the properties of the final networks, while the solid lines account for the evolution -- whenever available -- of those properties throughout the growth process. The graphs have been built by averaging uniform random distributions of $N=100$ points considering $N_{real} = 50$ different realizations.}
\label{fig:eglob_and_eloc_vs_ltot}
\end{figure}

Starting from a random distribution of nodes in a unit square, we built the corresponding MST, GT, EEM, and GN networks and computed their $\Eglob$ and $\Eloc$.  The results are shown in Fig.~\ref{fig:eglob_and_eloc_vs_ltot} and denote remarkable differences amidst the benchmarks. Preliminary work confirmed that differences persist when one considers different spatial distributions of points, that is, they do not depend critically on the details of the node positions. Additionally, the analysis of spatial layouts of the empirical networks considered in this study (see Sec.~\ref{sec:application}), showed that the distribution of distances among the points is compatible with a random uniform distribution.

%
%
%
%

In Fig.~\ref{fig:eglob_and_eloc_vs_ltot}, points correspond to the networks grown until their total length is almost the same as the GT one. Lines, instead, account for the intermediate stages corresponding to the growth phase -- \ie progressive addition of edges -- of the model (if available). A first glance at the panels reveals some interesting features. The first one is the opposite behavior of EEM and GT, with EEM performing better than GT in terms of $\Eglob$, and the other way around for $\Eloc$. Another feature is the fact that MST has a nonzero value of $\Eglob$ but $\Eloc = 0$, although this is expected since the removal of a single node in a tree implies the impossibility of communicating between its neighbors. Finally, the position of GN networks indicates higher values of both efficiencies but at a higher total cost (\ie total link length). This is due to the fact that GN networks are built using a cost function different from the mere spatial one.

The analysis of Fig.~\ref{fig:eglob_and_eloc_vs_ltot} highlights how differently the benchmark models behave with respect to the two efficiencies. Such differences can be leveraged and used to characterize each network using both $\Eglob$ and $\Eloc$. Thus, we can represent each network with the pair of values $(\Eloc ,\Eglob )$, which corresponds to a point in a two dimensional $[0,1]\times [0,1]$ diagram.

Any topology lies inside this square, its position depending on its specific features. For instance, a set of isolated nodes (networks with no links) will lie at the lower left corner $(0,0)$, while the upper right one $(1,1)$ corresponds to the complete graph which, by definition, has the maximum possible efficiency. Topologies having $(\Eloc, 0)$ or $(\Eloc, 1) \; \forall \Eloc \in \; ]0,1[ $ are not allowed since $\Eglob=0$ and $\Eglob=1$ can be obtained exclusively by a set of isolated nodes or a complete graph, respectively. Topologies falling on the $\Eloc = 0$ line correspond to tree-like (acyclic) graphs, while those falling on the $\Eloc = 1$ line are ensembles of disconnected complete subgraphs.

\begin{figure}
\centering
\includegraphics[width=\columnwidth]{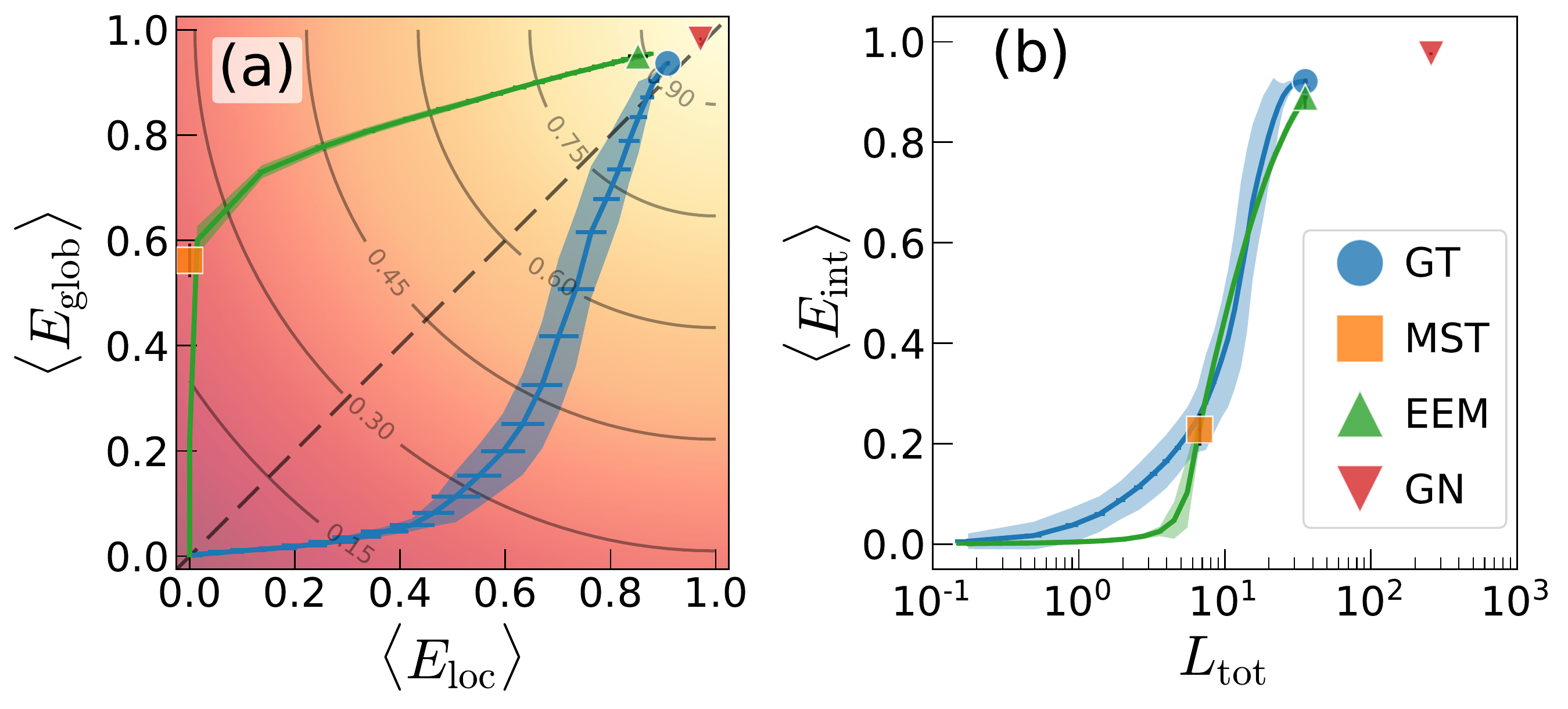}
\caption{(panel a) Comparison between the averages local, $\avg{\Eloc}$, and global, $\avg{\Eglob}$, efficiencies for the GT, MST, EEM, and GN topologies. The values of the other parameters are the same of those used in Fig.~\ref{fig:eglob_and_eloc_vs_ltot}.  Continuous black lines stand for $\Eint$ equilines. (panel b)  Average integrated efficiency, $\avg{\Eint}$, as a function of the total length of the system, $\Ltot$.}
\label{fig:eI_vs_ltot_and_eglob_vs_eloc}
\end{figure}

Every model of link growth -- \ie a model that builds networks by progressively adding connections to a set of initially isolated nodes -- draws a trajectory in the diagram starting from $(0,0)$ and, if not bounded to stop earlier, reaching $(1,1)$ (see Figs.\,\ref{fig:eglob_and_eloc_vs_ltot} and \ref{fig:eI_vs_ltot_and_eglob_vs_eloc}a).
In this sense, we can regard each real network as an intermediate stage of an unknown growth model, ideally connecting the point $(0,0)$ to $(1,1)$. Such a framework provides us with a metric to assess directly how efficient a given topology is from an overall viewpoint: the normalized distance between the point representing the considered topology and the upper right corner of the diagram (\ie the final target of any network growth model).

Therefore, we adopt this metric, which we call \emph{integrated efficiency}, as a comprehensive measure of the efficiency of spatial networks:
\begin{equation}
\label{eq:integ_eff}
\Eint = 1 - \sqrt{\frac{(1-\Eglob )^2+(1- \Eloc )^2}{2}} \,.
\end{equation}
which is the Euclidean distance to the $(1,1)$ coordinates in the two-dimensional space displayed in Fig.~\ref{fig:eI_vs_ltot_and_eglob_vs_eloc}(a). The above definition satisfies a crucial general consideration about the efficiency of real spatial networks: They perform reasonably well at both local and global scale \cite{banavar-prl-2000,katifori-prl-2010}. Indeed, this specific formulation encapsulates equally both scales by rewarding the balance between the two efficiencies. Consider the alternative, much simpler, measure $\Eint^\prime=(\Eglob +\Eloc )/2$. Three hypothetical “topologies” located at coordinates $(0,1)$, $(1,0)$, and $(0.5,0.5)$, respectively, would score the same in terms of $\Eint^\prime$. On the contrary, the proposed measure $\Eint$ takes a higher value in the third case, enhancing the balance between $\Eglob$ and $\Eloc$. The behavior of $\Eint$ for benchmark models as a function of $\Ltot$ is displayed in Fig.~\ref{fig:eI_vs_ltot_and_eglob_vs_eloc}b.

For simplicity, we consider a formulation of $\Eint$ where the contributions of $\Eloc$ and $\Eglob$ are equal. However, it is possible to alter Eq.~\eqref{eq:integ_eff} including a tunable multiplicative factor to account for asymmetric contributions. 

%
%
%
%

\section{Comparing network designs}
\label{sec:assessing}

The measure introduced in the previous section informs whether a certain topology is more, or less, efficient than another. Our final goal, however, is to compare the design of spatial networks in terms of resource allocation, something that is conceptually different. When we compute a network's integrated efficiency, we are calculating -- by definition -- how far it lies from the complete graph, which is the only graph reaching maximum values of $\Eloc$ and $\Eglob$ simultaneously. Nonetheless, there is a limit to how close a system can get to such an extreme, which is conditioned by the total cost $\Ltot$. A simple solution to this issue would be to divide the integrated efficiency, $\Eint$, by $L_{\text{tot}} / L_{\text{cg}}$, where $L_{\text{cg}}$ is the cost of the complete graph with the same node layout. However, such a rescaling procedure implies assuming a linear dependence of the efficiency on the total link length which in general is not true. For the purpose of a fair comparison, it is essential to avoid arbitrary assumptions that may introduce biases.

\subsection{Quasi-exact comparisons. A numerical recipe.}
\label{ssec:opt_model}

Our proposal is based on a very simple idea: to build the best possible network for a given amount of resources (\ie a given total cost $\Ltot$) and fixed node layout. In this manner, for every network under study, we can construct its optimal counterpart and hence compute $\Delta \Eint = \Eint^{\text{opt}} - \Eint $, where $\Eint$ is the integrated efficiency of the original network and $\Eint^{\text{opt}}$ that of its optimal counterpart. This value quantifies the room for improvement in a design with the same amount of resources. More importantly, if we consider two systems with different total cost and spatial scale, $\Delta \Eint$ enable to perform an indirect but fair comparison between them. 

Our goal is not to build an improved version of real networks -- which should display all their ideal characteristics -- but, rather, to  generate appropriate benchmarks by computing the maximum value that a given object function could take under certain constraints. 

\begin{figure}
\centering
\includegraphics[width=0.95\columnwidth]{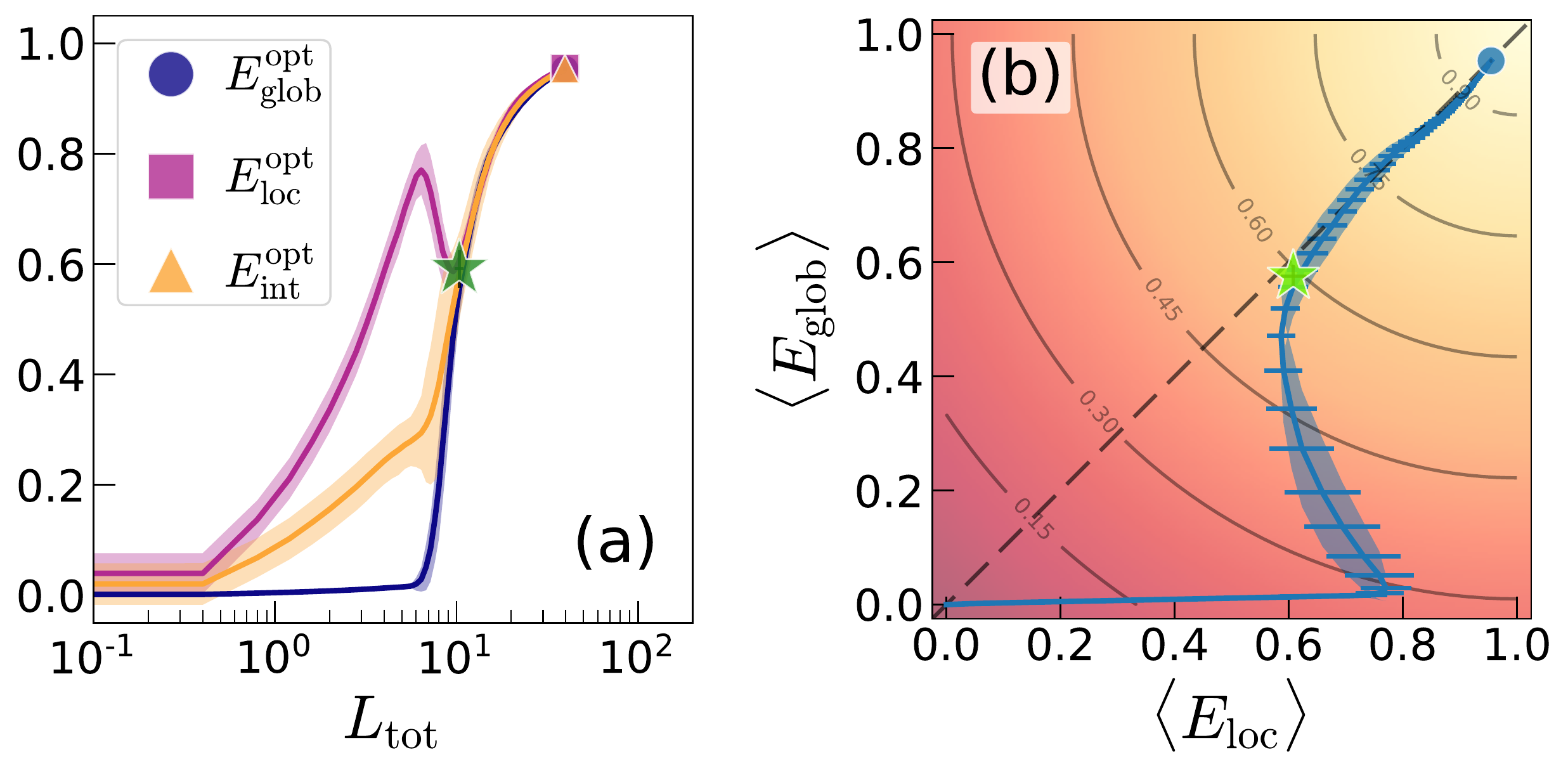}
\caption{(panel a) Evolution of the local ($\Eloc^{\text{opt}}$), global ($\Eglob^{\text{opt}}$), and integrated ($\Eint^{\text{opt}}$) efficiencies with respect to the total length of the system, $\Ltot$, for the networks generated using our optimal model. (panel b) Evolution of the average local ($\avg{\Eloc}$), and global ($\avg{\Eglob}$) efficiencies for the same networks. The green star denotes the values at which 90\% of the nodes belong to the giant component.  Results are averages over $N_{real} = 100$ different layouts of $N=100$ nodes uniformly distributed on the unit square. Continuous black lines stand for $\Eint$ equilines.}
\label{fig:eI_vs_ltot_and_eglob_vs_eloc_models}
\end{figure}

As a consequence, we have designed a growth algorithm that works as follows: we start considering an empty graph $G$ with $N$ nodes. Then, we add edges iteratively until the total length of the graph reaches the desired one. 
Adding edges means adding consecutive line segments in the $(L,\Eint)$ plane (see Figs.~\ref{fig:eI_vs_ltot_and_eglob_vs_eloc_models}\,a and \ref{fig:e_i-vs-l_tot-opt_mod-rescaling-size}\,a), connecting the origin with the  $(\Ltot,\Eint(\Ltot))$ point. Since we want to maximize the value of $\Eint(L)$ for $L=\Ltot$, at each iteration we have to look for the segment with the maximum vertical slope, \ie the edge maximizing the ratio between the variation of $\Eint$ and the increase in total cost:
\begin{equation}
\label{eq:target_function}
\max \left\{\frac{\tilde{E}_{\text{int}} - \Eint}{\tilde{L}-L}\right\}= \max_{\{i,j\}}\left\{\frac{\tilde{E}_{\text{int}} - \Eint}{d_{ij}}\right\} \,,
\end{equation}
where $L$ and $\Eint$ are the total weight of links and the integrated efficiency at the current step, respectively, and $\tilde{L}$ and $\tilde{E}_{\text{int}}$ stand for the same quantities after adding the link $(i,j)$. Since the identification of the edge to add involves the evaluation of the contribution of all the possible candidates, the overall procedure is completely deterministic.

Even though it is not possible to ensure that the topologies produced by such an algorithm reach the maximum possible value of the integrated efficiency, there are strong hints that they are very close to it. The optimization of $\Eint$ is a non-Markovian process and there exists the chance that different choices, locally not optimal, could lead to a better final result. To address this issue, we have explored the possibility to use alternative search methods based on simulated annealing. Our conclusion was that slightly higher values of $\Eint$ may be reached by a limited number of alternative topologies -- perhaps with a different balance between the local and global efficiencies --, requiring a considerably higher computational cost (\ie exploring the space of the configurations more exhaustively) in order to be discovered\footnote{Therefore, it is technically a \emph{quasi-optimal} algorithm but we avoid that nomenclature for the sake of clarity.}.

\indent In Fig.~\ref{fig:eI_vs_ltot_and_eglob_vs_eloc_models}, we display the average behavior of $\Eglob$, $\Eloc$, and $\Eint$ against $\Ltot$ for the networks generated using our optimal algorithm. We also report the evolution of the global and local efficiencies across the growth process according to the bi-dimensional representation adopted in Fig.~\ref{fig:eI_vs_ltot_and_eglob_vs_eloc}a. In particular, averages are computed over one hundred random distributions of $N=100$ nodes within the unit square. Panel (a) of Fig.~\ref{fig:eI_vs_ltot_and_eglob_vs_eloc_models} tells us that the algorithm first favors increases in $\Eloc$ up to a point where it is not possible to increase $\Eint$ further at the expense of $\Eloc$.  Unlike $\Eglob$, this metric has a non-monotonous behavior. For example, a system made of separated cliques has a $\Eloc=1$, and adding any other link will reduce $\Eloc$. This is the evolution's least interesting phase since, at this stage, systems are mainly composed by many connected components while the overall connectivity is extremely low. After the peak, the algorithm begins to link these isolated components at the expense of $\Eloc$  increasing $\Eglob$. When the total cost is roughly equal to that of the MST, the curves of $\Eloc$, $\Eglob$, and $\Eint$ merge together and start to behave in the same way.

%
%
%
%

\subsection{Approximated comparisons. An estimated universal upper bound.}
\label{ssec:norm_length}

\begin{figure}
\centering
\includegraphics[width=\columnwidth]{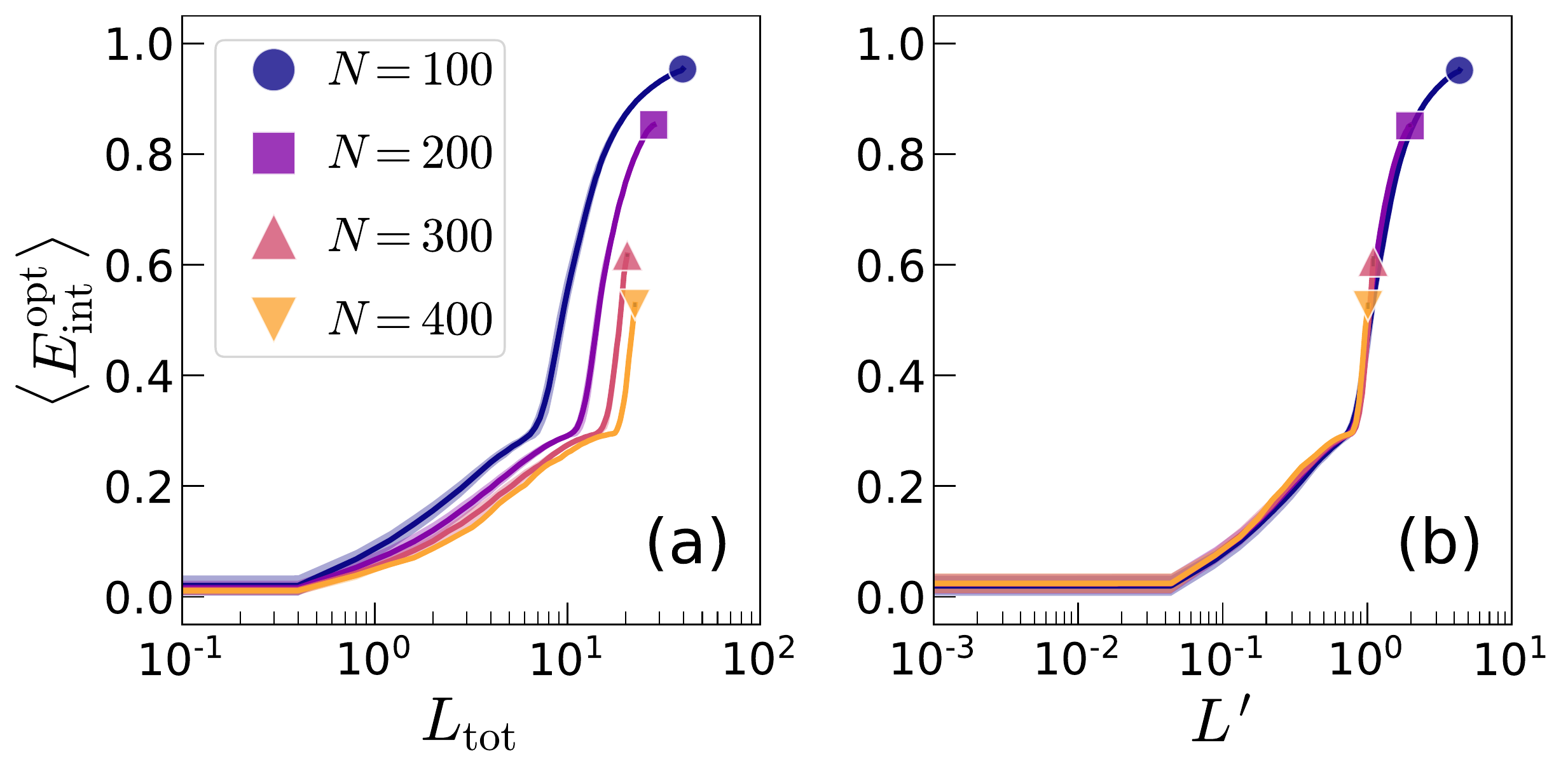}
\caption{(panel a) Evolution of the average integrated efficiency, $\avg{\Eint^{\text{opt}}}$, with respect to the total length $\Ltot$ for systems with different number of nodes $N$.  Solid lines account for the evolving systems, while symbols denote the final networks. (panel b) Same quantity of panel a, but displayed as a function of the rescaled length $\Lnorm$. The curves corresponding to different network sizes overlap perfectly.}
\label{fig:e_i-vs-l_tot-opt_mod-rescaling-size}
\end{figure}

The above algorithm provides an optimal network counterpart usable as a reference. However, in practice, the increase of the system size $N$ severely affects the runtime of the algorithm -- in particular of the calculation of $\Eloc$, -- jeopardizing the applicability of our methodology to large size systems. To overcome this limitation, we determined the expected value of $\Eint$ for any value of $\Ltot$ and any $N$, in the case of layouts of nodes randomly distributed in a square. First, we studied the behavior of the integrated efficiency against the overall cost, for several layouts of $N \in \{ 200, 300, 400 \}$ nodes. Figure \ref{fig:e_i-vs-l_tot-opt_mod-rescaling-size}a displays the behavior of the $\avg{\Eint}$ curves, confirming that the results are robust across sizes. Specifically, the increase in size translates into a shift of the $\Eint$ curve towards higher values of $\Ltot$.

By rescaling the $x$ coordinate of plots in Fig.~\ref{fig:e_i-vs-l_tot-opt_mod-rescaling-size}, we collapse them into a single, ``universal'' one which is the same for any value of $N$. This leads to the definition of a \emph{normalized total cost} of a network $G$, $L^\prime$, which reads: 
\begin{equation}
\label{eq:norm_length}
\Lnorm = \frac{\Ltot/ \avg{d} }{\left( \Lcg / \avg{d}  \right) ^{\alpha }} \,,
\end{equation}
where $\avg{d}$ stands for the average spatial distance among nodes and $\Lcg$ is the length of the complete graph having the same node layout as $G$. Such a normalized length can be rewritten as a combination of two variables: $\avg{d}$ and $N$, since the length of a complete graph is \linebreak $ \Lcg =\avg{d} \cdot \frac{N(N-1)}{2} \approx \avg{d}  \frac{N^2}{2}$. Hence, we obtain:
\begin{equation}
\label{eq:norm_length_def}
\Lnorm \simeq \frac{\Ltot}{\avg{d}  \cdot N^{2\alpha }} \,.
\end{equation}
We have found that $\alpha =1 / 3$.
As we can observe in Fig.~\ref{fig:e_i-vs-l_tot-opt_mod-rescaling-size}b, the rescaling of $\Ltot$ returns perfectly overlapped curves. Such a new, universal, curve allows us to compute the expected maximum value of $\Eint$ for a system with a given $\Lnorm$, providing an approximate upped bound to perform an indirect comparison between different systems. Specifically, given an empirical network with a certain $\Lnorm$, we can use the difference between its actual value of integrated efficiency and its expected maximum value, as a proxy of the system's performance. 

To improve the usability of this upper bound in real-world applications, it would be extremely useful to provide an analytical expression for the dependence of the expected maximum value of $\Eint$ on $\Lnorm$. However, for $L^\prime <1$, networks are usually disconnected and the behavior of the optimal integrated efficiency is very noisy (see also  Fig.~\ref{fig:eI_vs_ltot_and_eglob_vs_eloc_models}a).
In order to avoid this difficulty, we restricted the range of admissible values of the normalized length to $\Lnorm \geq 0.91$ (thus discarding the region where the slope of the curve is almost zero) and fitted our numerical data to the relation:

\begin{equation}
\Bar{E}_{\mathrm{int}}^{\mathrm{opt}}(L^\prime) =1 - c_1(L^\prime-c_2)^{-c3} \,.
\label{fit}
\end{equation}

Using non-linear least squares, we found that $c_1= 0.187$, $c_2 = 0.406$, and $c_3 = 1.211$. In this way, given any real network, it is possible to calculate the expected upper bound for its integrated efficiency without generating any artificial network, but simply from the average distance between its nodes and its total cost, through $\Lnorm$ and Eq.~\eqref{fit}. Including shorter total link lengths (\ie $L^\prime \leq 0.91$) would need a separate specific discussion which goes beyond the scope of the present work.

%
%
%

\section{Applications}
\label{sec:application}

To illustrate the applicability of the proposed analytical tools, here we use them to compare some real spatial networks based on their integrated efficiency. Specifically, we consider seven different collections of networks, a few of them (UK-Flights, Cities, and Latium Vetus/Southern Etruria) correspond to successive snapshots of the same evolving system. The collections are:
\begin{description}
 \item[UK Flights] Time-varying network of domestic flights in the United Kingdom between years 1990 and 2003 \cite{uk-airlines}. Nodes correspond to airports, while an edge between two airports accounts for the distance among them. For each year/graph, we keep only those routes with, at least, 5000 carried passengers across the year.
 \item[Cities] The evolution of urban street patterns of a small region in northern Italy, captured in four snapshots between 1955 and 2007 \cite{strano-scirep-2012}. Nodes correspond to the intersection between two streets or dead ends, while the weight of an edge corresponds to the length of the street connecting two nodes. For computational reasons, we consider only a small -- rectangular -- sample of the whole dataset centered around a single village.
\item[Latium Vetus/Southern Etruria] The networks of trails among settlements between 950 and 509 BC (Iron Age) in two regions of Italy, namely: Latium Vetus (LV) \cite{fulminante-front-2017} and Southern Etruria (SE) \cite{prignano-jas-2019}. Nodes represent settlements, while an edge denotes a direct route connecting them. From older to earlier, we have five snapshots for both regions: Early Iron Age 1 Early , Early Iron Age 1 Late, Early Iron Age 2, Orientalizing Age and Archaic Age. We label these snapshots chronologically as (LV$x$/SE$x$) with $x \in [1,5]$. The snapshots do not have the same duration, but the properties of the system are more or less ``stable'' within each snapshot.
 \item[Catalonia railway] This network describes the current regional railway network in Catalonia \cite{in-preparation1}. Nodes correspond to aggregated groups of contiguous towns, while edges denote the length of the railway line connecting them.
 \item[Hispania roads] The networks of trails among cities and towns in Hispania (Iberian peninsula) during the Roman Empire \cite{in-preparation2}. As for the Latium Vetus and Etruria collections, nodes represent settlements, while an edge denotes a direct route connecting them.
 \item[Rome railway] The network of rail connections in Rome, where nodes represent stops/stations and link constitutes a direct connection between two nodes \cite{kujala-scidata-2018}. Weights correspond to the geodesic distance between both ends of the link. The original data-set splits many stops in two, each one corresponding to the two ways of the line. We simplify the network by merging stops having the same name into a single node.
 \item[Power grid] A simplified model of the power grid network of Italy, where transmission lines are assumed bidirectional and identical. It is a subset of a model of the continental European electricity system, comprising 1494 buses and 2156 lines. \cite{power_dataset}.
 
\end{description}

The main topological features of all these networks are reported in Tab.~\ref{tab:dataset}. Such table reveals the diversity of the networks under analysis.

\subsection{Direct comparison of efficiencies}

Efficiency values in Tab.~\ref{tab:dataset} shows interesting features among the variety of datasets. First, we see that  $\Eloc$ and $\Eglob$ in the UK-Flights fall within a narrow range of values centered around $0.6$, despite the edge density $\rho$ is fairly high. On the other side, we noticed that all the rest of networks are all \emph{terrestrial}, and tend to be more efficient at a global rather than local scale. This is in line with the principles behind the design and growth of such kind of networks, which tend to privilege tree-like structures spanning the whole system at the expenses of resilience \cite{banavar-prl-2000,barthelemy-jstat-2006,bohn-prl-2007,katifori-prl-2010}. In this sense, terrestrial infrastructure networks are likely to show fairly high $\Eglob$, since they provide paths among all nodes with little chance to large route factors. Nevertheless, exceptions are found. For example, the rail network of the city of Rome shows two connected components, dragging down the value of $\Eglob$ compared to other similar systems. On the other end, terrestrial networks are more vulnerable at the local level, as denoted by their values of $\Eloc$.

%
%
%
\begin{table*}
\begin{center}
\begin{scriptsize} 
\begin{tabular}[c]{m{1.7cm}|*{2}{c} d| *{3}{d} | d | *{3}{d} |c}  
\cline{2-12}
   & $N$ & $K$ &  \multicolumn{1}{c}{$\rho$} & \multicolumn{1}{|c}{$\Ltot$} & \multicolumn{1}{c}{$\Lcg$} & \multicolumn{1}{c|}{$\avg{d}$} & \multicolumn{1}{c|}{$L^\prime$} & \multicolumn{1}{c}{$\Eloc $} & \multicolumn{1}{c}{$\Eglob $} & \multicolumn{1}{c|}{$\Eint $} & Ref.\\
  & & & \multicolumn{1}{c|}{$(\%)$} & \multicolumn{3}{c|}{$(Km)$ or $(m)$} & \multicolumn{1}{c|}{$\quad$} & \multicolumn{3}{c|}{$\quad$} & \\ \hline
\textbf{UK Flights} &  \multicolumn{10}{l|}{$\quad$} & \cite{uk-airlines} \\ \hline
1992 & 41 & 130 & 15.85 & 44833.310 & 328656.403 & 400.800 & 12.149 & 0.663 & 0.636 & 0.649 &  \\
1995 & 39 & 141 & 19.03 & 50215.194 & 304749.131 & 411.267 & 13.729 & 0.688 & 0.680 & 0.684 &  \\
1998 & 41 & 146 & 17.80 & 54019.694 & 333391.976 & 406.576 & 14.431 & 0.664 & 0.669 & 0.666 & \\
2001 & 39 & 141 & 19.03 & 52949.882 & 305950.002 & 412.888 & 14.419 & 0.553 & 0.614 & 0.582 & \\
\hline
\textbf{Cities} &  \multicolumn{10}{l|}{$\quad$} & \cite{strano-scirep-2012} \\ \hline
1955 & 29 & 36 & 8.87 & 4474.270 & 148386.739 & 365.485 & 1.692 & 0.343 & 0.819 & 0.518 &  \\
1980 & 71 & 98 & 3.94 & 12383.492 & 1417732.535 & 570.516 & 1.618 & 0.407 & 0.824 & 0.563 &  \\
1994 & 80 & 110 & 3.48 & 13111.289 & 1763976.734 & 572.534 & 1.587 & 0.432 & 0.818 & 0.579 &  \\
2007 & 90 & 124 & 3.10 & 14152.991 & 2385978.985 & 609.290 & 1.485 & 0.423 & 0.819 & 0.572 &  \\ \hline
\textbf{Latium\newline Vetus} & \multicolumn{10}{l|}{$\quad$} & \cite{fulminante-front-2017} \\ \hline
LV1 & 93 & 198 & 4.63 & 1359.614 & 111083.691 & 25.966 & 3.249 & 0.678 & 0.890 & 0.760 &  \\
LV2 & 93 & 198 & 4.63 & 1318.020 & 108710.242 & 25.411 & 3.218 & 0.648 & 0.875 & 0.736 &  \\
LV3 & 107 & 239 & 4.21 & 1625.535 & 143007.035 & 25.217 & 3.637 & 0.679 & 0.871 & 0.755 &  \\
LV4 & 93 & 220 & 5.14 & 1637.569 & 102742.846 & 24.017 & 4.231 & 0.687 & 0.876 & 0.762 &  \\
LV5 & 78 & 189 & 6.29 & 1471.613 & 75232.370 & 25.052 & 4.107 & 0.731 & 0.917 & 0.801 &  \\ \hline
\textbf{Southern\newline Etruria} & \multicolumn{10}{l|}{$\quad$} & \cite{prignano-jas-2019} \\ \hline
SE1 & 116 & 199 & 2.98 & 1586.445 & 271499.897 & 40.705 & 2.082 & 0.507 & 0.875 & 0.641 &  \\
SE2 & 115 & 207 & 3.16 & 1660.375 & 265460.422 & 40.497 & 2.204 & 0.571 & 0.887 & 0.686 &  \\
SE3 & 130 & 235 & 2.80 & 1811.973 & 344291.037 & 41.060 & 2.183 & 0.606 & 0.869 & 0.706 &  \\
SE4  & 168 & 311 & 2.22 & 2062.108 & 605568.932  & 43.169  & 1.989  & 0.851  & 0.649  & 0.730  & \\
SE5  & 179 & 366 & 2.30 & 2300.959  & 679017.117  & 42.622  & 2.153  & 0.868 & 0.714 & 0.777 & \\ 
\hline
\textbf{Catalonia} & \multicolumn{10}{l|}{$\quad$} & \cite{in-preparation1} \\ \hline
Railway & 34 & 37 & 6.60 & 1040.580 & 54481.616 & 97.115 & 1.299 & 0.233 & 0.637 & 0.400 & \\
\hline
\textbf{Hispania} & \multicolumn{10}{l|}{$\quad$} & \cite{in-preparation2} \\ \hline
Roads & 89 & 127 & 3.24 & 10115.003 & 1742076.283 & 444.861 & 1.453 & 0.460 & 0.812 & 0.595 & \\ \hline
\textbf{Rome} & \multicolumn{10}{l|}{$\quad$} & \cite{kujala-scidata-2018}\\ \hline
Railway & 80 & 103 & 3.26 & 377.297 & 39333.959 & 12.447 & 2.066 & 0.246 & 0.538 & 0.375 & \\ \hline
\textbf{Power grid} & \multicolumn{10}{l|}{$\quad$} &  \cite{power_dataset} \\ \hline
Italy & 139 & 207 & 2.16 & 11914.776 & 4087233.378 & 426.153 & 1.316 & 0.467 & 0.794 & 0.596 & \\
\hline
\end{tabular}
\end{scriptsize} 
\caption{Summary of the topological indicators for all the empirical datasets. For each network, we report its number of nodes, $N$, of edges, $K$, the edge density, $\rho$, the total length for the empirical, $\Ltot$, and complete graph, $\Lcg $, as well as the average spatial distance among the nodes, $\avg{d}$, the rescaled length of the system, $L^\prime$, and their local $\Eloc $, global $\Eglob $, and integrated $\Eint $ efficiencies. Finally, for each network dataset, we report the bibliographic source of the data.}
\label{tab:dataset}
\end{center}
\end{table*}

Beyond comparing the efficiencies of the different spatial networks in a descriptive way, our methodology allows for a systematic comparison of diverse spatial networks considering both their $\Eint$ and $L_{tot}$.

\subsection{Systematic comparison considering total lenght}

For each empirical network, we generate an optimized one using the algorithm presented in Sec.~\ref{sec:assessing} while preserving the node layout (\ie their position) and the total length, $\Ltot$. 

\begin{figure}
\centering
\includegraphics[width=\columnwidth]{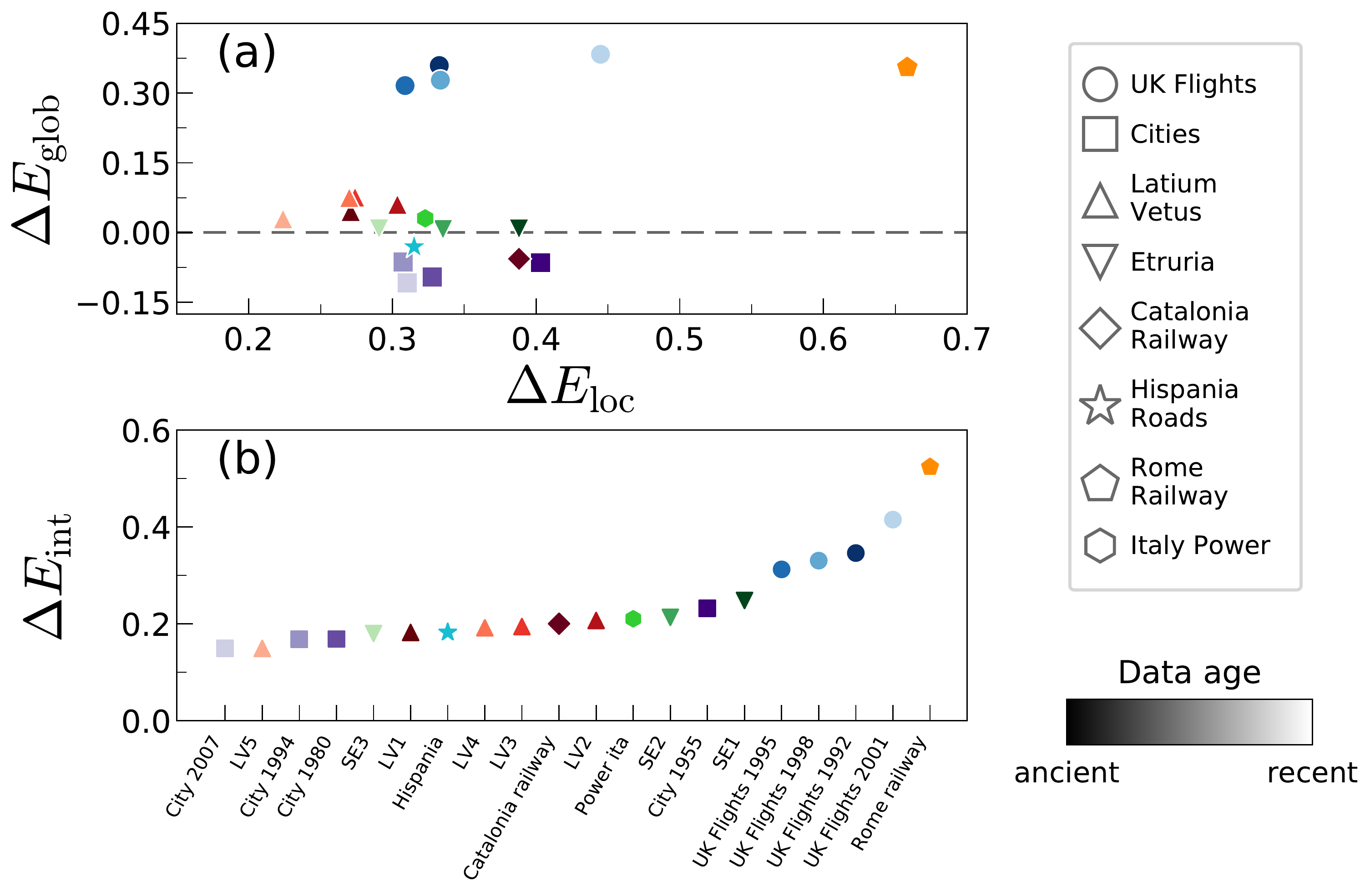}  
\caption{(panel a) Global efficiency difference, $\Delta \Eglob = E^{\text{opt}}_{\text{glob}} - \Eglob$ , with respect to the same quantity calculated for the local efficiency, $\Delta \Eloc $ for all the dataset reported in Tab.~\ref{tab:dataset}. The hue of the color is used to order the dataset in a chronological way. (panel b) Ranking of the empirical systems according to the integrated efficiency difference $\Delta \Eint = \Eint^{\text{opt}} - \Eint$.  Points for data SE4 and SE5 are not shown since, as mentioned in Sec.~\ref{ssec:norm_length}, we were unable to generate the optimal counterparts.}
\label{fig:empirical_data}
\end{figure}

First, we analyse the differences on both local and global efficiencies separately. We compute the differences $\Delta E_X = E^{\text{opt}}_X - E_X\,, \; X \in \{ loc, glob \}$ among the efficiencies of the empirical and optimal networks. In Fig.~\ref{fig:empirical_data}a we report the values of $\Delta \Eloc $ and $\Delta \Eglob $ for all the cases under scrutiny. It is worth noting that $\Delta \Eloc $ is always negative, while this is not the case of $\Delta \Eglob  $. With the exception of UK-Flights and Rome Railway, all the other collections tend to have values of $\Eglob $  close to the optimal counterpart one (\ie $\Delta \Eglob = 0$). A closer inspection to $\Delta\Eglob$ highlights interesting features. One is that for Cities, Hispania, and Catalonia Railway networks $\Delta\Eglob < 0$ (\ie they are more efficient than their optimized counterparts). Latium Vetus, Etruria and the Italian Power Grid, instead, fall very close to the optimum. Another interesting feature is that UK-Flights and Rome Railway networks are sub optimal both locally and globally, confirming our guess about the existence of criteria behind their design other than merely spatial ones.

We sort all the datasets according to their $\Delta \Eint = \Eint^{\text{opt}} - \Eint$ and check how far the real networks lie from the upper bound (Fig.~\ref{fig:empirical_data}b). The extent of efficiency's difference tells us how much better the systems could have performed consuming the same amount of resources. The differences range from $\Delta \Eint \approx 15\%$  (City 2007, LV5) to above $50\% $ (Rome Railway), while the majority of the real networks are about $20\% $ less efficient than their optimal counterparts.

%
%
%
%

\begin{figure}
\centering
\includegraphics[width=\columnwidth]{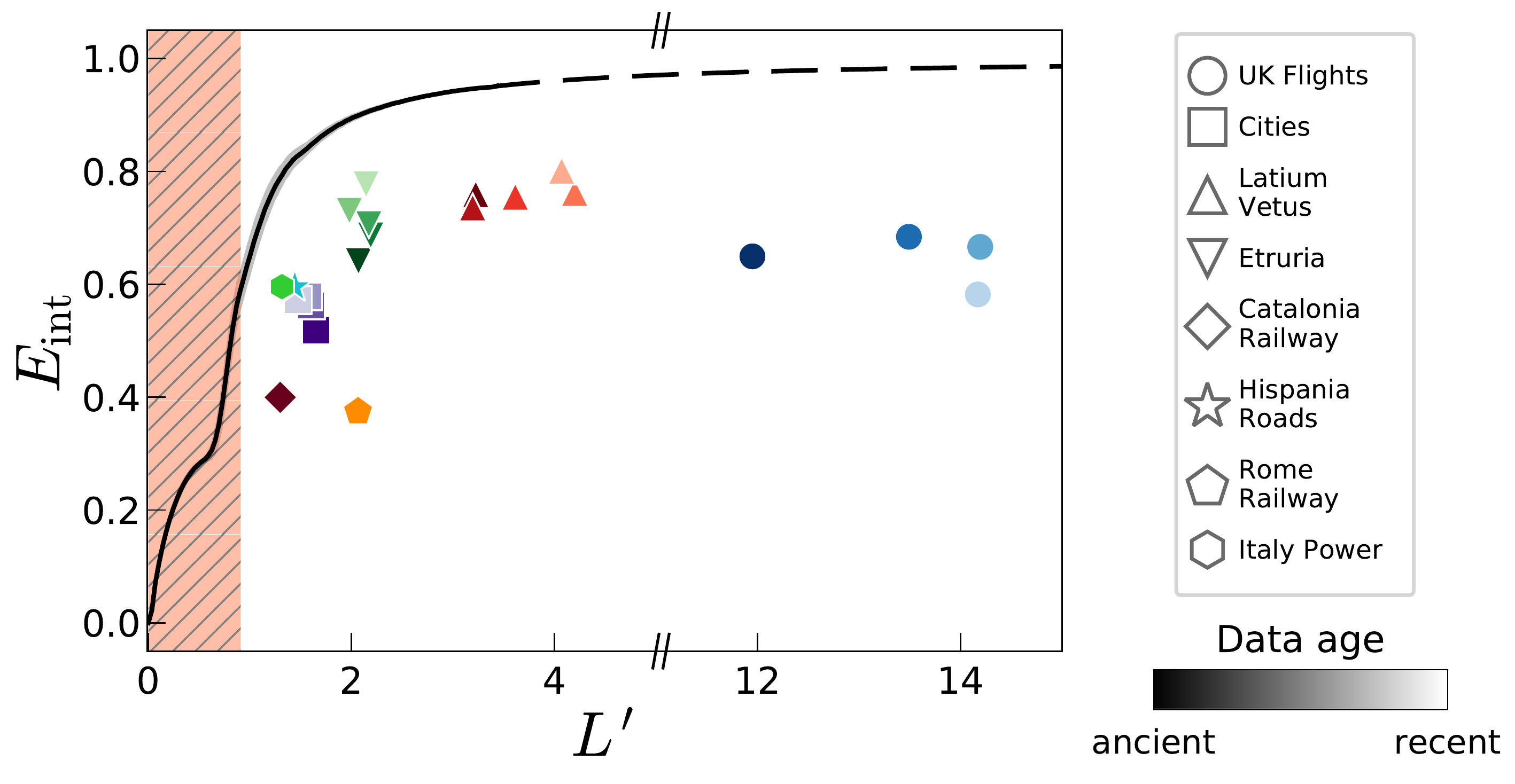}
\caption{Integrated efficiency, $\Eint$ as a function of the rescaled length $L^\prime$ for the same datasets. The filled area for $L^\prime \leq 0.91$ denotes the region for which the system has not fully percolated yet. The hue of the color is used to order the dataset in a chronological way. The continuous black line stand for the optimal curve  $\Eint^{\text{opt}}(L')$.}
\label{fig:empirical_data_2}
\end{figure}

We have thus proven that it is relevant to consider the upper bound of the integrated efficiency of each real network since it provides novel, complementary information with respect to the mere value of $\Eint$. However, as discussed in Sec.~\ref{ssec:norm_length}, the computational cost of determining such upper bound increases rapidly with the size and total cost of the system under scrutiny. Therefore, it is interesting to assess whether replacing the value of $\Eint^{\text{opt}}$ with its expected value provided by Eq.~\eqref{fit} leads to similar results. By doing so, we are disregarding the details of the node layouts, while still considering their overall characteristics through the average node distance. In Fig.~\ref{fig:empirical_data_2} we report the values of the empirical $\Eint$ as a function of the rescaled length $L^{\prime }$. As expected, all the values of $\Eint$ in empirical systems lie way below the optimal curve. Generally speaking, networks with higher connectivity are more efficient than those with less links, albeit being more costly. However, this is not always true. By looking at Fig.~\ref{fig:empirical_data_2} we notice that UK Flights networks attain approximately the same efficiency of Southern Etruria ones, but with a much higher cost.\newline
\indent We rank networks according to $\Delta \Eint^\prime = \Bar{E}_{\mathrm{int}}^{\mathrm{opt}} - \Eint$, the difference between their $\Eint$ and the corresponding value computed through Eq.~\eqref{fit}. We find that rankings based on $\Delta \Eint^\prime$ and on $\Delta \Eint$ provide a Spearman brank correlation of $r_s =0.618$ ($\text{p-val}=0.0037$). This indicates that the curve is a valid alternative to ranking networks according to the actual difference between optimal and empirical integrated efficiencies. Nevertheless, the correlation's value highlights non negligible discrepancies between the rankings. The reason for such differences is that real spatial layouts differ from layouts in our simulations. Random distribution of nodes in a square show very little fluctuations as indicated by the shadow of the curve. On the contrary, a real network's layout can be far from this distribution, thus affecting the output of a spatial network model \cite{pablo_marti-scirep-2017}. 

%
%
%

\subsection{Application to networks' comparison in meaningful scenarios}

We have shown that our methodology can be used with diverse spatial networks in a plethora of application domains. Now, we aim at presenting how such comparison can be used to address relevant research questions. In the following, we use some of the empirical networks presented above to highlight two meaningful ways of applying our methodology. The first example is an archaeological case of study based on the Latium Vetus and Southern Etruria datasets \cite{fulminante-front-2017, prignano-jas-2019}. These two data-sets describe the pathway networks connecting settlements in two neighboring regions with similar characteristics through time. Since there are obvious differences in terms of covered distance, system size and total cost (length) among the ten datasets, the networks corresponding to the two regions cannot be compared directly. On the contrary, our methodology allows assessing which region hosted better designed infrastructures by looking at the efficiency difference to their optimal counterparts. A second example would involve analysing the evolution of roads in the Cities' dataset. There, we can assess the efficiency in the area and the evolution of a system that increases both in size and total cost, but for which we initially cannot know how better could have performed under those constraints. Despite the strong changes in structural terms (\eg size increase and density decrease), our methodology would allow for a longitudinal comparison throughout the temporal snapshots in the data-set. Even when the computation of the optimal is jeopardized by the system size (like in 1994 and 2007) we could use the upper bound curve to obtain their estimation.

\section{Conclusions}

This manuscript provides tools to compare the design of different spatial networks in terms of the duality \emph{resources employed} and \emph{quality reached}. Following an approach similar to \cite{zamora-commphys-2019}, such a comparison is performed indirectly by contrasting the quality improvement that each one of the networks' designs could reach. 

The paper presents the different components of our tool-set progressively. First, we have introduced the notion of \emph{integrated efficiency}, $\Eint$, as a metric to quantify spatial networks performance at global and local scale simultaneously, while rewarding the balance between the two. Second, we propose an algorithm to computationally estimate the upper bound of our quality function for a given specific network: we have devised an algorithm to generate networks with maximal $\Eint$ ($\Eint^{\mathrm{opt}}$) with the same node layout and total cost as in the original network. The smaller is the difference between such an upper bound and the empirical value, the higher we consider the design quality of the network under analysis to be. 

Since the high computational cost of the optimal network algorithm may hinder its applicability on large networks, we provide a universal expression for approximated upper bound to any network. Considering a setting of $N$ nodes randomly distributed in a unit square, we computed the expected maximal value of $\Eint$ ($\Bar{E}_{\mathrm{int}}^{\mathrm{opt}}$) as a function of the total cost $\Ltot$. Then, by defining a rescaled total link length, $L^\prime$, we successfully collapsed the $\Bar{E}_{\mathrm{int}}^{\mathrm{opt}}$ versus $\Ltot$ curves for different sizes onto a single one. In this way, we have been able to express $\Bar{E}_{\mathrm{int}}^{\mathrm{opt}}$ as a function of the number of nodes $N$, the average distance between nodes $\langle d \rangle$, and the total cost $\Ltot$ of the network under study.

Finally, to test the applicability of our method, we have analyzed the performance of a heterogeneous set of spatial networked systems. We have checked that our approach provides new information beyond the mere comparison between two networks' efficiency. We provide also several examples of application among the included datasets, showing how it helps to compare networks with different size and total length in a meaningful way.

In conclusion, we have shown that a meaningful comparison of spatial networks cannot be exempt from the definition of proper upper bounds with specific cost constraints. This can be done (almost) exactly, by running our maximal efficiency algorithm, or approximately, thanks to the universal curve. The latter constitutes a good approximation for systems whose size does not make the computation of $\Eint^{\text{opt}}$ feasible. However, the particularities of the layout (especially for low $\Lnorm$) may affect the precision of the method. In the future, it will be worth exploring how the specificities of a layout affects a systems' $\Eint^{\text{opt}}$ with respect to the value provided by the curve. Another direction to pursue is to explore the effect of weighting differently the contributions of the local and global efficiencies in the computation of $\Eint$.

\section*{Data availability}
The following datasets are available at \cite{data_repo}: Latium Vetus, Southern Etruria, Catalonia railway and Hispania roads. The rest of the datasets are available online in the following references: Rome railway \cite{kujala-scidata-2018} and Power grid \cite{power_dataset}. Regarding Cities datasets, please contact the author (see Acknowledgements section). The code used for running the algorithm can also be found in \cite{data_repo}.


\begin{acknowledgments} 
The authors thank Simon Carrignon for his help with computational issues, and Michael Gastner for helping with the implementation of the GN model. The authors also thank Emanuele Strano, Francesca Fulminante, Lluc Font-Pomarol, Pau de Soto, Tom Brughmans and Sergi Valverde for providing the data on evolving cities, archeological networks, and the power grid.\\
 
AC acknowledges the support of the Spanish \emph{Ministerio de Ciencia e Innovacion} (MICINN) through Grant IJCI-2017-34300. AC and SL acknowledge the financial support of \emph{Ministerio de Economia y Competitividad} (MINECO) through grant RYC-2012-01043. IM, LP and AD-G acknowledge the financial support of the \emph{European Research Council} within the Advanced Grant EPNet (340828). AD-G acknowledges  financial  support  from  FIS2015-71582-C2-2 -P (MINECO/FEDER), and the Generalitat de Catalunya (Project No.2014SGR-608).\\
\end{acknowledgments} 

IM, AC, AD-G, LP and SL designed research and wrote the article; IM and AC wrote code and performed simulations; IM analyzed numerical results.

\appendix

\section{The Gastner-Newman model}
\label{sec:gn_model}

We present here a brief description of the Gastner-Newman (GN) model to generate spatial networks introduced in \cite{gastner-epjb-2006}. The main idea behind the model is that there are two types of ``costs'' associated with a given network: one related with the construction of the infrastructure, and another related with its usage. Given a graph with $N$ nodes and $K$ edges, $G(N,K)$ \cite{boccaletti-phys_rep-2006}, we consider its embedding in the bidimensional space, $\mathbb{R}^2$. We denote with $d_{ij}$ the Euclidean distance between nodes $i$ and $j$, respectively. Therefore, the total \emph{cost of construction} of graph $G$, $T$, reads:
\begin{equation}
\label{eq:gn-costr-cost}
T = \sum_{i=1}^N \sum_{j=i+1}^N a_{ij} \, d_{ij}\,.
\end{equation}
Where $a_{ij}$ is the element of the adjacency matrix, $\cal{A}$, of the graph \cite{latora-book-2017}. The total \emph{usage cost}, $Z_\lambda$, instead, is:
\begin{equation}
\label{eq:gn-user-cost}
Z_\lambda = \sum_{i=1}^N \sum_{j=i+1}^N \tilde{l}_{ij}(\lambda) \,,
\end{equation}
with $\tilde{l}_{ij}$ being the \emph{shortest path length} between nodes $i$ and $j$ which, in turn, is the sum of the lengths of the edges forming the \emph{path} between $i$ and $j$ \cite{latora-book-2017}. The path length depends on a parameter $\lambda$ such that:
\begin{equation}
\label{eq:dist-funct}
\tilde{l}_{ij}(\lambda)  = \lambda \sqrt{N} \, d_{ij} + (1 - \lambda) \,,
\end{equation}
with $0 \leq \lambda \leq 1$. Parameter $\lambda$ accounts for the users' perception of distances. For $\lambda = 0$, users give more importance to paths made of few hops, while for $\lambda = 1$ they pay more attention to shorter paths (in terms of distance). Finally, the total cost, $C$, of the whole graph is:
\begin{equation}
\label{eq:gn-total-cost}
C(G) = T + \gamma Z_\lambda \,,
\end{equation}
where parameter $\gamma \in [0,1]$ determines the relative weight between construction and usage cost. The GN algorithm generates \emph{optimal} spatial networks minimizing cost $C$. The cost optimization can be implemented using either greedy or simulated annealing techniques \cite{newman-book-2012}. We decided to implement the latter, since it ensures higher probabilities of finding the optimal network. In our case, we were interested in building GN networks with a given total cost. Hence, given a spatial network $G^\star$, we compute its cost $C^\star$ according to Eq.~\eqref{eq:gn-total-cost}. Considering the same nodes layout of $G^\star$, we build a GN network $G$ through the following steps.
\begin{enumerate}
 \item Create the complete graph $G_0$, and compute its cost $C(0)$.
 \item{At each step, $t$, perform with equal probability one of these two operations:%
 \begin{enumerate}
  \item Add/Remove an edge:\newline
  Choose a random pair of nodes $i$ and $j$, and if they are connected (\ie $\exists \, e_{ij}$) we remove the corresponding edge. Otherwise, we add the edge. The removal can take place unless one of the two nodes has degree one (\ie otherwise the node will get disconnected).
  \item Rewiring:\newline
  Choose an edge $e_{ij}$ at random. Then, choose a node $k \neq {i,j}$ at random and create the edge $(i,k)$ or $(j,k)$ -- if it does not exist already. Finally, remove the edge $e_{ij}$.
 \end{enumerate}
 } 
 \item{Compute the cost of the resulting graph, $C(t)$, and then:%
 \begin{enumerate}
  \item Accept the move with the following probability $p$:%
  \begin{equation}
  \nonumber
  p =
  \begin{cases}
  \exp\Bigl[ -\beta \bigl( C(t) - C(t-1)\bigr) \Bigr]  & \text{ if } C(t) > C(t-1)\\
  1 & \text{ otherwise }
  \end{cases}
  \end{equation}
  \item increase the value of $\beta$ of a quantity $\Delta \beta = 1 +  3 \cdot 10^{-6}$ (with $\beta(0) = \tfrac{0.1}{C_{MST}}$, and being $C_{MST}$ the cost of the minimum spanning tree of the nodes layout).
 \end{enumerate}
 } 
 \item Repeat stages 2 and 3 either $t_\infty = 1.5\times10^6$ times, or until $C \simeq C^\star$.
\end{enumerate}

\section{The Greedy Triangulation}
\label{sec:gt_model}

Here we provide a brief description on how to compute the \emph{Greedy Triangulation} graph of a given layout of nodes embedded into a bidimensional metric space $\mathbb{R}^2$. Given a set of $N$ nodes embedded into a two dimensional space, the most connected (planar) triangulation graph $G^P$ has -- at most -- $K=3N-6$ edges \cite{bondy-book-2008}; and the maximally connected triangulation minimizing its total length, $L$, is called the \emph{minimum weighted triangulation} (MWT). Since no polynomial time algorithm is known to compute the MWT, following \cite{buhl-epjb-2004,cardillo-pre-2006} we consider -- instead -- its \emph{greedy} approximation known as Greedy Triangulation (GT).
\begin{figure}[b]
\centering
\includegraphics[width=0.6\columnwidth]{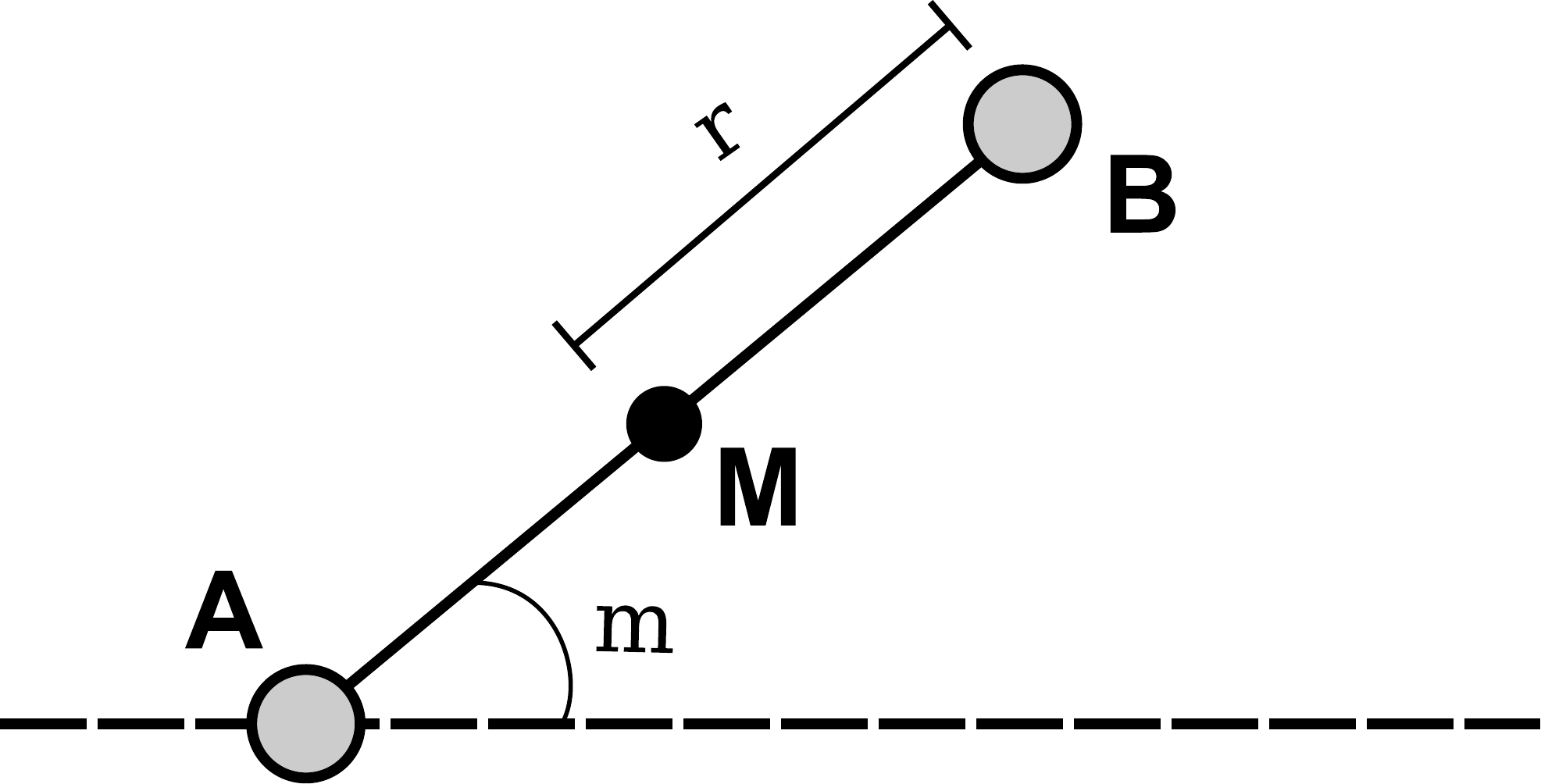}
\caption{Schematic representation of a segment $\overline{AB}$.}
\label{fig:segment}
\end{figure}
Taking $N$ points embedded into a bidimensional Euclidean space, we compute first the corresponding complete weighted graph $G_0$ with $K=\tfrac{N(N-1)}{2}$ edges. Then, we prepare a list of $G_0$'s edges, sorted in ascending order of length (\eg using quicksort \cite{cormen-book-2001}). Given the schematization displayed in Fig.~\ref{fig:segment}, for each edge/segment, $\overline{AB}$, connecting nodes $A,B \in G_0$ we store the following information:
\begin{equation}
\nonumber \left\lbrace \text{ID}_A, x_A, y_A, x_M, y_M, \text{ID}_B, x_B, y_B, r, m \right\rbrace \,.
\end{equation}
Where $\text{ID}_i, x_i, y_i$ are the ID, and $x,y$ coordinates of node/point $i$, $M$ is the middle point of $\overline{AB}$, $r = \tfrac{d_{AB}}{2}$ is its semi-length, and $m$ is its angular coefficient (\ie the tangent of the angle between $\overline{AB}$ and the $x$-axis). In its essence, the algorithm to compute the GT graph, $G_{GT}$, parses the sorted edgelist of $G_0$ and checks whether each candidate edge $e^\star \in G_0$ belongs to the GT or not. After adding the shortest edge/segment of $G_0$ to the empty set of edges of $G_{GT}$, we check if any other edge of $G_0$'s edge list intersects (and eventually how) with those of $G_{GT}$. To check if two segments $\overline{AB} \in G_{GT}$ and $\overline{CD} \in G_0$ intersect, -- and assuming that $x_A \leq x_B$ and $x_C \leq x_D$, -- we compute the distance between their middle points $d_{12} = d_{M_{\overline{AB}} \, M_{\overline{CD}}}$. Then:
\begin{enumerate}
 \item If $d_{12} > (r_{AB} + r_{CD})$ the two segments do not intersect, and thus $\overline{CD}$ potentially belongs to $G_{GT}$.
 \item If $d_{12} \leq (r_{AB} + r_{CD})$ the two segments may intersect, and we need to perform further checks to include/exclude the edge from $G_{GT}$.
\end{enumerate}
To perform such checks, we have to compute the coordinates of the intersection point, $(X,Y)$, which are:
\begin{equation}
\label{eq:crossing_coord}
\begin{split}
X &= \dfrac{y_1 - m_{AB} \; x_1 - \left( y_2 - m_{CD} \; x_2\right)}{m_{CD} - m_{AB}}\, , \\
Y &= \dfrac{m_{CD} \; y_1 - m_{AB} \; y_2 + m_{AB} \; m_{CD} \left( x_2 - x_1\right)}{m_{CD} - m_{AB}} \, .
\end{split}
\end{equation}
Where $x_\alpha, y_\alpha \; \alpha \in \{1,2\}$ are the coordinates of the middle point of $\overline{AB}$ if $\alpha = 1$, or $\overline{CD}$ if $\alpha = 2$. If $X \in [\min(x_A, x_C), \max(x_B, x_D)]$, then the two segments will cross for sure (since $d_{12} \leq (r_{AB} + r_{CD})$) and $\overline{CD}$ could be discarded (a similar criterion could be established for $Y$). For each candidate edge $\overline{CD}$, we repeat the procedure described above for all the edges $\overline{AB}$ already present in  $G_{GT}$.\\
\indent However, there are some exceptions to the ``intersection'' rule. In particular, segments sharing one vertex do ``technically'' intersect, but without breaking $G_{GT}$'s planarity and, hence, might belong to $G_{GT}$. Another case requiring special attention is that of segments either parallel to one axis or perpendicular to them. In such case the check on either $X$ or $Y$ alone is not enough, and we must ensure that both $X$ and $Y$ fall outside their respective intervals, instead. Lastly, for parallel segments (\ie $m_{CD} = m_{AB}$), the relations in Eqs.~\eqref{eq:crossing_coord} have a singularity and, thus, cannot be used to compute the coordinates of the intersection point. If segment $\overline{CD}$ does not cross any of those of $G_{GT}$, it can be added to $G_{GT}$ and we proceed to check the next candidate of $G_0$'s edgelist. The algorithm stops either when $3N-6$ edges have been added to $G_{GT}$, or if no more candidates to check are available. In the latter case, the number of edges of the GT will be lesser than $3N-6$. This is due to the fact that some edges between nodes laying at the outskirt of the node layout might be added without breaking the planarity. However, such edges cannot be represented as straight lines, and thus their intersection cannot be computed using the above mentioned method. The amount of missing edges is approximately in the order of $\sqrt{N} \ll 3N - 6$.

\section{Equitable Efficiency Model}
\label{sec:ee_model}

In this section, we present the essential traits of the Equitable Efficiency Model (EEM) introduced by Prignano \etal in \cite{prignano-jas-2019}. EEM is a growth model which builds networks from empty and static spatial node layouts. In its essence, the model adds one link at a time ensuring that such addition constitutes the best improvement in the efficiency of communication among any pair of nodes. 

Given a node layout embedded in a two dimensional space, $\mathbb{R}^2$, at each step we calculate the route factor, $E_{ij}$, (\ie the ratio between the spatial distance, $d_{ij}$, and the shortest path length, $l_{ij}$) between all the pair of nodes $i$ and $j$. According to its definition, $E_{ij} \in [0,1] \, \forall \, i,j$; where $E_{ij} = 0$ when $i$ and $j$ belong to different components of the system (\ie $l_{ij} = \infty$), and $E_{ij} = 1$ when they are directly connected, instead. After computing all the values of $E_{ij}$, we sort them in ascending order. The connection having the smallest $E_{ij}$ is added to the network, and the above procedure is repeated iteratively until the graph has a total length, $\Ltot$, equal to the desired one. However it is worth noting that, according to the definition of route factor, $E_{ij} = 0$ for all nodes belonging to different components, regardless of their distance. To ensure a parsimonious usage of resources, and avoid an arbitrary selection of one of the unconnected pairs, we ideally replace $l_{ij} = \lim_{\Lambda\to\infty}\Lambda$ with $l_{ij} = \Lambda$ where $\Lambda \gg \Lcg $ is a large, but finite, length. This replacement implies that the route factor between pairs of nodes belonging to different components will be ranked according to their spatial distance. Therefore, until the graph has one single component, the algorithm will select connections between unreachable nodes, starting from those that are physically closer to each other. This means that the set of links connecting the nodes into a single component is nothing else than the Minimum Spanning Tree (MST) of the layout under consideration.

\end{document}